\documentclass[twocolumn,showpacs,preprintnumbers,amsmath,amssymb]{revtex4}
\usepackage{graphicx}
\usepackage{dcolumn}
\usepackage{bm}
\usepackage{color}
\usepackage{bbold}

\begin{document}

\title{Photon-antibunching in the fluorescence of statistical ensembles of emitters at an optical nanofiber-tip} 
\author{Elmer Suarez$^1$}
\author{David Auw\"arter$^1$}
\author{Tiago J. Arruda$^2$}
\author{Romain Bachelard$^3$}
\author{Philippe W. Courteille$^2$}
\author{Claus Zimmermann$^1$}
\author{Sebastian Slama$^1$}

\affiliation{$^1$ Physikalisches Institut and Center for Collective Quantum Phenomena in LISA$^+$, Eberhard Karls Universit\"at T\"ubingen, Auf der Morgenstelle 14, D-72076 T\"ubingen, Germany\\
$^2$ Instituto de Fisica de Sao Carlos, Universidade de Sao Paulo, 13566-590 Sao Carlos, Sao Paulo, Brazil\\
$^3$ Departamento de Fisica, Universidade Federal de Sao Carlos, 13565-905 Sao Carlos, Sao Paulo, Brazil}

\date{\today}

\begin{abstract}
This proposal investigates the photon-statistics of light emitted by a statistical ensemble of cold atoms excited by the near-field of an optical nanofiber tip. Dipole-dipole interactions of atoms at such short distance from each other suppress the simultaneous emission of more than one photon and lead to antibunching of photons. We consider a mean atom number on the order of one and deal with a poissonian mixture of one and two atoms including dipole-dipole interactions and collective decay. Time tracks of the atomic states are simulated in quantum Monte Carlo simulations from which the $g^{(2)}$-photon autocorrelation function is derived. The general results can be applied to any statistical ensemble of emitters that are interacting by dipole-dipole interactions.

\end{abstract}


\maketitle
\section{Introduction}
Single-photon sources are essential building blocks of devices in quantum information science \cite{Eisaman11}. An ideal source emits exactly one photon into a single optical mode at an arbitrarily chosen time, with each photon being indistinguishable from all others generated by the source. The demand that no more than a single photon is emitted at a time typically requires the use of single emitters, for instance single atoms, molecules, quantum dots, or single defects in crystal lattices \cite{Schubert92, Kuhn02,Hijlkema07,Lounis00,Nothaft12,Michler00,Senellart17,Schlehahn18,Lohrmann16}. The fabrication of such single emitter sources is however technically demanding \cite{Aharonovich16}, which motivates the search for new approaches that are simpler to build. The idea presented in this paper is to collect the fluorescence from a nanoscale volume that contains a statistically varying number of $N$ emitters with an average on the order of $\left<N\right>\sim 1$. This contains the possibility that more than one emitter is present within the volume. However, in this case the excitation and emission process can be dramatically altered due to interactions between the emitters. By using cold atoms as emitters the Doppler effect of moving atoms can be eliminated. We theoretically analyze the light statistics of the fluorescence of a statistical mixture of one and two atoms with and without interactions by simulating coherent excitation by a laser field emerging from the fiber tip, and incoherent emission using a quantum Monte-Carlo approach. From the resulting time stamps at which photons are emitted the corresponding $g^{(2)}$-function is calculated. The paper is organized as follows: Section II defines the problem of how statistical ensembles of emitters can emit light with subpoissonian statistics and large brightness, which at first glance seem to be contradicting requirements. Section III contains the theoretical model used for calculating the fluorescence from two atoms interacting with a classical pump light field and by dipole-dipole interactions among them. In Section IV an analytic solution of the Hamilton-operator (neglecting the decay terms) shows that an increasing interaction strength leads to a retardation of the simultaneous excitation of both atoms which is eventually responsible for the evolution of a subpoissonian photon statistics. Section V contains results of the quantum Monte-Carlo simulations with interaction strength $\delta_{12}$ and collective correction $\gamma_{12}$ to the decay rate as free parameters. The corresponding $g^{(2)}$ photon-autocorrelation functions exhibit photon-antibunching for increasing values of $\delta_{12}$.  Eventually, Section VI deals with the experimental situation of Rubidium atoms at the apex of a nanoscale fiber tip. The simulations take both the interatomic interaction and the interaction with the surface of the nanotip into account.  The dipole-dipole interaction strength $\delta_{12}$ and the collective correction $\gamma_{12}$ are determined using realistic parameters. The corresponding $g^{(2)}$ function is calculated from a large number of random atomic positions.\\ 
\begin{figure}[ht]
	\centerline{\scalebox{.9}{\includegraphics{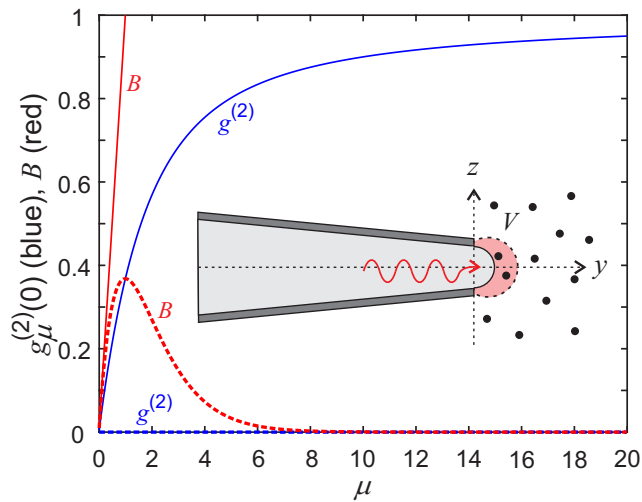}}}
	\caption{Sketch: a nanotip is immersed into a gas of cold atoms. The tip emits light and excites the fluorescence of a statistical number of atoms in a volume $V$ at the tip apex. Plots: without interactions, the corresponding value of $g^{(2)}_{\mu}(0)$ (blue solid line) is increasing with average atom number $\mu$ in the volume. The corresponding brightness $B$ of the source (red solid line) is proportional to $\mu$. If interactions suppress the emission from more than one atom, the value of $g^{(2)}_{\mu}(0)=0$ for any $\mu$ (blue dashed line), while the brightness exhibits a maximum (red dashed line).}
	\label{fig:tip_g2}
\end{figure}

\section{Light statistics of ensembles} 
The physical mechanism described in this paper can be applied to any statistical ensemble of emitters in a nanoscale volume. However, the specific system which we have in mind is that of an optical fiber which is tapered down at one of its ends to a nanotip in order to increase interactions with dipole emitters nearby the tip \cite{Chang07,Auwaerter13,Mihaljevic14,Arruda17}. We assume the tip is immersed into a gas of cold atoms with number density $\rho$, and a classical light field emerging from the fiber end excites atoms in a small volume $V$ in front of its apex, Fig.~\ref{fig:tip_g2}. Thus, an average number of 
\begin{equation}
\mu=\rho V
\label{eq1}
\end{equation}
atoms is within the volume. The exact number $N$ of atoms in the volume is poissonianly distributed. i.e. the probability of finding $N$ atoms in the volume is given by
\begin{equation}
P_\mu(N)=\frac{\mu ^{N}}{N!}e^{-\mu }. 
\label{eq:poisson}
\end{equation}
We will first show how the light statistics is influenced by such a poissonianly distributed number of non-interacting emitters. The $g^{(2)}(\tau)$-function 
\begin{equation}
g^{(2)}(\tau)=\frac{\left<\hat{a}^\dagger(t)\hat{a}^\dagger(t+\tau)\hat{a}(t+\tau)\hat{a}(t)\right>}{\left<\hat{a}^\dagger(t)\hat{a}(t)\right>^2}
\label{eq:g2_function}
\end{equation}
is proportional to the probability density of the emission of a photon at a time difference $\tau$ after the emission of a first photon, where the total photon annihilation operater $\hat{a}$ is defined by the annihilation operators $\hat{a}_j$ of all contributing atoms via
\begin{equation}
\hat{a}=\sum_{j=1}^{N}\hat{a}_j~.
\label{eq:annihil_operator}
\end{equation}
We assume that the interatomic distance is well below the optical wavelength such that propogation effects between photons emitted from distant atoms can be neglected. The $g^{(2)}$-function is scaled such that $g^{(2)}(\tau)\rightarrow 1$ for sufficiently large $\tau$ where no correlations between emission processes exist. Antibunching, which is the signature of single photon emission, corresponds to a value of $g^{(2)}(0)=0$. For a single atom the emission is antibunched due to the time it takes to re-excite the atom once it is projected to its ground state after the emission of a photon. For an arbitrary fixed number of $N$ independent atoms the value of $g^{(2)}_N(0)$ scales with the atom number as
\begin{equation}
g^{(2)}_N(\tau =0)=\frac{N-1}{N}.
\label{g2_N}
\end{equation}
This result is caused by the effect that, whenever one of the $N$ atoms emits a photon, any of the other $N-1$ atoms can emit a photon, too. Finally, the contributions of different atom numbers to the $g^{(2)}$-function are weighted with their statistical probability $P_\mu(N)$ of finding $N$ atoms in the volume and with the rate $R_N$ at which photons are emitted by the $N$ atoms: 
 \begin{equation}
 g^{(2)}_{\mu}(\tau =0)=\frac{1}{R}\sum_{N=1}^{\infty }g^{(2)}_N(\tau =0) P_\mu(N) R_N,
 \label{eq:g2_N_mean}
 \end{equation}
with normalization $R=\sum_N P(N) R_N$. Please note that $g^{(2)}$ - functions of different realizations can be averaged with their relative weight, when the timescale on which the realizations are stable is larger than the correlation time; in our case i.e. the timescale on which the position of an atom changes substantially. The condition is very well fulfilled for ultracold atoms at the nanotip. With a velocity of $1~\mathrm{mm/s}$ they move over a distance of $1~\mathrm{nm}$ within a time of $1~\mu\mathrm{s}$, which is much larger than the correlation time which is on a timescale of $1/\left(2\pi\gamma_0\right)=27~\mathrm{ns}$ for Rubidium with natural linewidth $\gamma_0$. For $N$ independent emitters the rate increases as $R_N=N R_1$, i.e. the number of emitted photons increases linearly with the number of atoms. Thus, the $g^{(2)}$-function quickly tends to a value of one for increasing mean atom number $\mu$, as shown in Fig.~\ref{fig:tip_g2}. Nevertheless, antibunching can be observed also with statistical ensembles of emitters, if $\mu \ll 1$, where -- in turn -- the probability $P(0)$ that no atom is in the volume gets high. This effect reduces the brightness $B$ of the source, similar to other non-deterministic sources like those based on parametric down-conversion \cite{Senellart17}. We define the brightness as total photon emission rate normalized to the photon emission rate of a single atom: 
 \begin{equation}
B(\mu)=\frac{1}{R_1}\sum_{N=1}^{\infty }P_\mu(N) R_N~. 
\label{eq:brightness}
\end{equation}
For independent atoms the brightness
 \begin{equation}
B(\mu)=\sum_{N=1}^{\infty }P(N) N=\mu
\label{eq:brightness_ind}
\end{equation}
is equal to the average atom number $\mu$, see Fig.~\ref{fig:tip_g2}. Interactions are able to reduce the emission rate $R_N$ for atom numbers $N>1$. In an extreme case, interactions would completely suppress the excitation of more than one atom, and the emission rate for a single atom would be $R_1=1$, whereas $R_N=0$ for $N>1$. The corresponding value of the correlation function would be $g^{(2)}(0)=0$ for any $\mu$ with a maximum brightness of $B\approx 0.37$ for $\mu=1$, see dashed lines in Fig.~\ref{fig:tip_g2}. The statistics of this single-photon source shows perfect antibunching while the brightness is comparable to a single-emitter based photon source. The following sections deal with the influence of dipole interactions on the light statistics. We focus specifically on the case of one and two atoms. This is justified by the fact that for increasing atom number in the volume the suppression will be even stronger due to the decreasing distance between the atoms. Moreover, for a mean atom number of $\mu=1$ the Poisson-probability that three atoms are within the volume is only $P(3)=0.06$.\\

\section{Quantum jump approach for two atoms}
\label{sec:QJA}
The theoretical method we use in this paper is known as the quantum jump approach to dissipative dynamics \cite{Dum92, Molmer93, Plenio98, Haroche06}. The atomic state $\left\vert \Psi_j \right>$ of atom number $j=1,2$ is described in the matrix basis
of the ground state $\left\vert g\right>$ and the excited state $\left\vert e\right>$
\begin{equation}
\left\vert g\right> =
\left( 
\begin{array}{c}
1 \\ 
0
\end{array}
\right),~ 
\left\vert e\right> =
\left( 
\begin{array}{c}
0 \\ 
1
\end{array}
\right)~.
\label{single_atom_states}
\end{equation}

The states are coupled by a laser field with detuning $\Delta=\omega_L-\omega_{0}$ between the laser frequency $\omega_L$ and the atomic transition frequency $\omega_{0}$ of the atoms. The Rabi frequency of the coupling is $\Omega_j=d \mathcal{E}_j /\hbar$, with dipole matrix element $d$ of the transition, and electric field amplitude $\mathcal{E}_j$ of the light at the position of the corresponding atom.  In the rotating frame, the effective single-atom Hamiltonian reads 
\begin{equation}
H_{j}=\hbar\left( 
\begin{array}{cc}
0 & \Omega_j /2 \\ 
\Omega_j /2 & \Delta
\end{array}
\right)~.
\label{eq:single_atom_hamiltonian} 
\end{equation}
The total wave function  $\left\vert \Psi \right\rangle$ describing the state of both atoms including correlations is an element of the four-dimensional tensor product Hilbert space with the vector basis
\begin{eqnarray}
\left\vert gg\right\rangle=\left\vert g\right\rangle \otimes \left\vert g\right\rangle &,& \left\vert ge\right\rangle =\left\vert g\right\rangle \otimes \left\vert e\right\rangle~, \nonumber \\
\left\vert eg\right\rangle=\left\vert e\right\rangle \otimes \left\vert g\right\rangle &,& \left\vert ee\right\rangle=\left\vert e\right\rangle \otimes \left\vert e\right\rangle~,
\end{eqnarray}
where the operator left and right to $\otimes$ corresponds to atom no. 1 and 2, respectively. The Hamiltonian acting on the tensor space is given by the single atom Hamiltonians (\ref{eq:single_atom_hamiltonian}) via
\begin{equation}
	H_\mathrm{a} =H_{1}\otimes \mathbb{1} +\mathbb{1}\otimes H_{2}~,
\end{equation}%
with two-dimensional identity matrix $\mathbb{1}$. Thus, the Hamiltonian without interactions is 
\begin{equation}
H_\mathrm{a}=\hbar\left( 
\begin{array}{cccc}
0 & \frac{\Omega_2}{2} & \frac{\Omega_1}{2} & 0 \\ 
\frac{\Omega_2}{2} & \Delta & 0 & \frac{\Omega_1}{2} \\ 
\frac{\Omega_1}{2} & 0 & \Delta & \frac{\Omega_2}{2} \\ 
0 & \frac{\Omega_1}{2} & \frac{\Omega_2}{2} & 2\Delta
\end{array}%
\right), 
\end{equation}
Dipole-dipole interactions between the two atoms are described by the interaction Hamiltonian \cite{Rudolph95} %
\begin{eqnarray}
W&=&\hbar\delta_{12}\left(S^+\otimes S^- + S^-\otimes S^+\right)\nonumber\\
&=&\hbar\left( 
\begin{array}{cccc}
0 & 0 & 0 & 0 \\ 
0 & 0 & \delta_{12} & 0 \\ 
0 & \delta_{12} & 0 & 0 \\ 
0 & 0 & 0 & 0%
\end{array}%
\right)~,
\label{eq:H_ia} 
\end{eqnarray}
with interaction potential strength $\delta_{12}$ and raising resp. lowering operators 
\begin{eqnarray}
S^+&=&\left\vert e\right\rangle \left\langle g\right\vert~,\\ 
S^-&=&\left\vert g\right\rangle \left\langle e\right\vert. 
\label{eq:loweringoperators} 
\end{eqnarray}
The complete Hamiltonian is thus given by 
\begin{equation}
H_\mathrm{tot}=H_\mathrm{a}+W~.
\label{eq:H_tot} 
\end{equation}
The evolution of the density matrix is determined by the master equation in its Lindblad form \cite{Rudolph95}
\begin{eqnarray}
\frac{d\rho}{dt}=&-&\frac{i}{\hbar}\left[H_\mathrm{tot},\rho\right]-\sum_{i,j=1}^2\gamma_{ij}\left(\frac{1}{2} S_i^+ S_j^- \rho \right. \nonumber \\
&+& \left. \frac{1}{2} \rho S_i^+ S_j^- - S_j^- \rho S_i^+ \right)~.
\label{eq:Lindblad} 
\end{eqnarray}
Here, the matrix
\begin{equation}
\gamma_{ij}=\left( 
\begin{array}{cc}
\gamma_1 & \gamma_{12} \\ 
\gamma_{12} & \gamma_2
\end{array}
\right)
\label{eq:gamma_ij} 
\end{equation}
is defined with decay rates $\gamma_1$ and $\gamma_2$ of the individual atoms and collective correction $\gamma_{12}$ to the decay rate which will be introduced in Section VI, and 
\begin{eqnarray}
S_1^{\pm}=S^{\pm}\otimes\mathbb{1}~,\\
S_2^{\pm}=\mathbb{1}\otimes S^{\pm}~.
\end{eqnarray}
The quantum state trajectory is simulated in time steps with length $dt$. For each time step the probability that a jump $m=1,2$ occurs is given by
\begin{equation}
	p_{m} = \left<\Psi\right|L_m^\dagger L_m \left|\Psi\right> dt~.
	\label{eq:jump_probabilities}
\end{equation}
The collective jump operators 
\begin{eqnarray}
L_1 &=& \sqrt{\Lambda_1}\left(\alpha S_1^- + \beta  S_2^-\right),\\
L_2 &=& \sqrt{\Lambda_2}\left(-\beta S_1^-+\alpha S_2^-\right)~. 
\label{eq:jump_operators}
\end{eqnarray}
are derived following \cite{Haroche06} as eigenvectors of the matrix (\ref{eq:gamma_ij}) with eigenvalues 
\begin{eqnarray}
\Lambda_1 &=& \bar\gamma+\sqrt{(\gamma_{12})^2+\Delta_\gamma^2}~, \\
\Lambda_2 &=& \bar\gamma-\sqrt{(\gamma_{12})^2+\Delta_\gamma^2}~, 
\label{eq:Lambdas}
\end{eqnarray}
where 
\begin{eqnarray}
\bar\gamma &=& \frac{1}{2}\left(\gamma_1+\gamma_2\right)~, \\
\Delta_\gamma &=& \frac{1}{2}\left(\gamma_1-\gamma_2\right)~.
\label{eq:gammabar}
\end{eqnarray}
They are linear combinations of the single atom jump operators $S_j^-$ with coefficients
\begin{eqnarray}
\alpha&=&\frac{\Delta_\gamma+\sqrt{(\gamma_{12})^2+\Delta_\gamma^2}}{\sqrt{(\gamma_{12})^2+\left(\Delta_\gamma+\sqrt{(\gamma_{12})^2+\Delta_\gamma^2}\right)^2}}~,\\
\beta&=&\frac{\gamma_{12}}{\sqrt{(\gamma_{12})^2+\left(\Delta_\gamma+\sqrt{(\gamma_{12})^2+\Delta_\gamma^2}\right)^2}}~.
\label{eq:alpha_beta}
\end{eqnarray}
For equal single atom decay rates $\gamma_1=\gamma_2=\gamma$, the collective jump operators are given by 
\begin{eqnarray}
L_1 &=& \sqrt{\frac{\gamma+\gamma_{12}}{2}}\left(S_1^- + S_2^-\right)~,\\
L_2 &=& \sqrt{\frac{\gamma-\gamma_{12}}{2}}\left(-S_1^- + S_2^-\right)~, 
\label{eq:jump_operators_2}
\end{eqnarray}
corresponding to superradiant and subradiant decay, respectively. In the case considered in Sec. VI where the atoms may have different distance from the nanotip, the single atom decay rates may also differ from each other. Finally, the decay terms can be included into an effective Hamiltonian
\begin{equation}
H_\mathrm{eff}=H_\mathrm{tot}-iJ~,
\label{eq:H_eff}
\end{equation}
with 
\begin{equation}
J=\frac{1}{2}\sum_m L_m^\dagger L_m~.
\label{eq:J_matrix}
\end{equation}
The calculated probabilities (\ref{eq:jump_probabilities}) for the individual jumps are compared with a random evenly distributed number $r\in\left[0,1\right]$.
If $r<p_0=1-(p_1+p_2)$ no jump occurs, and the wavefunction evolves according to Hamiltonian dynamics with $H_{\textrm{eff}}$. As the effective Hamiltonian is not Hermitian (due to the decay terms), the wavefunction is renormalized to its absolute value after each step. The Hamiltonian dynamics can thus be summarized as 
\begin{equation}
\left\vert \Psi \right\rangle \rightarrow \left\vert \Psi \right\rangle^{\prime }=\frac{\exp (-iH_{\textrm{eff}}dt)\left\vert \Psi \right\rangle }{\left\vert \exp (-iH_{\textrm{eff}}dt)\left\vert \Psi \right\rangle \right\vert ^{2}}. 
\end{equation}
Note that the value of $dt$ has to be chosen such that $p_1+p_2\ll 1$.\\
If, on the other hand, a jump occurs, i.e. for $r>p_0$, which jump operator $L_m$ is applied is given by the value of $r$ compared to the smallest integer $m=1,2$ such that $\sum_{n=0}^{m}p_n>r$. This condition is a clever way of determining the jump operator depending on its probability. After the jump, the wavefunction is renormalized like in the Hermitian case. Thus, the jump dynamics is summarized as 
 \begin{equation}
\left\vert \Psi \right\rangle \rightarrow \left\vert \Psi \right\rangle^{\prime }=\frac{L_m\left\vert \Psi \right\rangle}{\left\vert L_m \left\vert\Psi\right\rangle \right\vert ^{2}}~. 
\end{equation}
Before we apply this quantum jump approach to specific situations we present in the following section a general argument why dipole interactions can lead to the suppression of a simultaneous emission of photons from both atoms. For that reason we analytically solve the Schr\"{o}dinger equation for two atoms including the dipole-dipole interactions.

\section{Analytical solution for two atoms in free space including dipole-dipole interactions}
The Schr\"{o}dinger equation with the Hamiltonian $H_{\text{tot}}$ from (\ref{eq:H_tot}) including dipole-dipole interactions (but without the decay terms) can be solved analytically. We consider here the case where $\Delta=0$, and $\Omega_1=\Omega_2=\Omega$ in order to keep the calculation as simple as possible. Thus, we have
\begin{equation}
H_{\text{tot}}=\hbar\left( 
\begin{array}{cccc}
0 & \frac{1}{2}\Omega  & \frac{1}{2}\Omega  & 0 \\ 
\frac{1}{2}\Omega  & 0 & \delta_{12} & \frac{1}{2}\Omega  \\ 
\frac{1}{2}\Omega  & \delta_{12} & 0 & \frac{1}{2}\Omega  \\ 
0 & \frac{1}{2}\Omega  & \frac{1}{2}\Omega  & 0%
\end{array}%
\right)
\label{eq:H_tot_simple} 
\end{equation}
The formal solution of the Schr\"odinger equation is
\begin{equation}
\vert\Psi(t)\rangle=e^{-\frac{i}{\hbar}H_{\text{tot}} t}\vert\Psi(0)\rangle ~.
\label{eq:solution} 
\end{equation}
The matrix exponential in (\ref{eq:solution}) can be solved analytically. If initially both atoms are in the ground state, i.e. $\Psi(0)=\left\vert gg\right\rangle$ the projection of $\Psi(t)$ onto the doubly excited state is  given by 
\begin{equation}
<ee|\Psi(t)>=\frac{\Omega^2}{4}\left(\frac{3\delta_{12}-A}{B}e^{\frac{i(\delta_{12}-A)t}{2}} + \frac{3\delta_{12}+A}{C}e^{\frac{i(\delta_{12}+A)t}{2}}\right) ~,
\label{eq:P_ee} 
\end{equation}
with parameters
\begin{eqnarray}
A&=&\sqrt{\delta_{12}^2+4\Omega^2} ~,\\
B&=&\frac{1}{2}\left(\delta_{12}-A\right)^3-\left(\delta_{12}^2+\Omega^2\right)\left(\delta_{12}-A\right)-\delta_{12}\Omega^2~,\\
C&=&\frac{1}{2}\left(\delta_{12}+A\right)^3-\left(\delta_{12}^2+\Omega^2\right)\left(\delta_{12}+A\right)-\delta_{12}\Omega^2~.
\label{eq:P_ee_params} 
\end{eqnarray}
\begin{figure}[ht]
	\centerline{\scalebox{.9}{\includegraphics{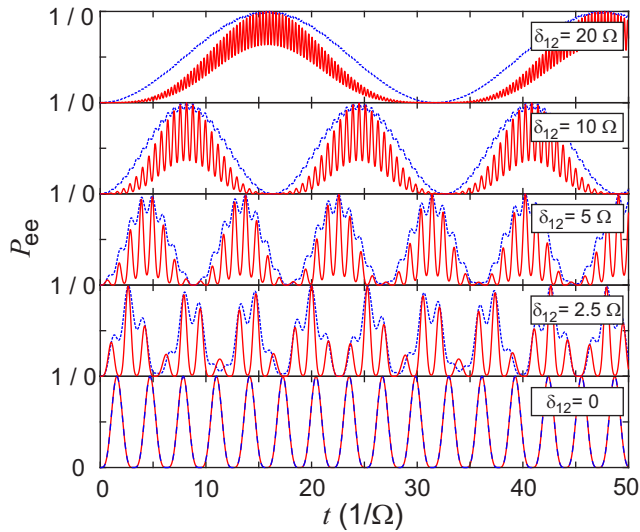}}}
	\caption{Time tracks of the doubly excited state occupation $P_\text{ee}(t)$ (red solid lines) for varying interaction strength. In these plots $\Delta=0$, and $\Omega_1=\Omega_2$. The stronger the interaction strength is, the longer it takes to excite both atoms. The product of the probabilities that respectively one of the atoms is excited (blue dashed lines) differs from $P_\text{ee}(t)$ for interaction strength $\delta_{12}\neq 0$.}
	\label{fig:Pee}
\end{figure}
The probability of a double-excitation $P_\mathrm{ee}(t)=\left|<ee|\Psi(t)>\right|^2$ is plotted in Fig.~\ref{fig:Pee} for varying interaction strength $\delta_{12}$. An increasing interaction strength slows down the timescale of the excitation process. This is the crucial feature for the suppression of the photon emission by more than one atom and leads to a suppression of the simultaneous emission of more than one photon. A further interesting feature can be observed in Fig.~\ref{fig:Pee}, when $P_\mathrm{ee}(t)$ is compared with the product of the probabilities that respectively one of the atoms is excited, whereas the state of the other atom is arbritrary. For $\delta_{12}\neq 0$ these probabilities differ from each other. This discrepancy is interpreted as interaction-induced entanglement between the two atoms and will be analyzed in a separate work. In the following section the decay term is reintroduced, and the quantum jump approach is applied in order to derive corresponding $g^{(2)}$-functions.

\begin{figure}[ht]
	\centerline{\scalebox{.9}{\includegraphics{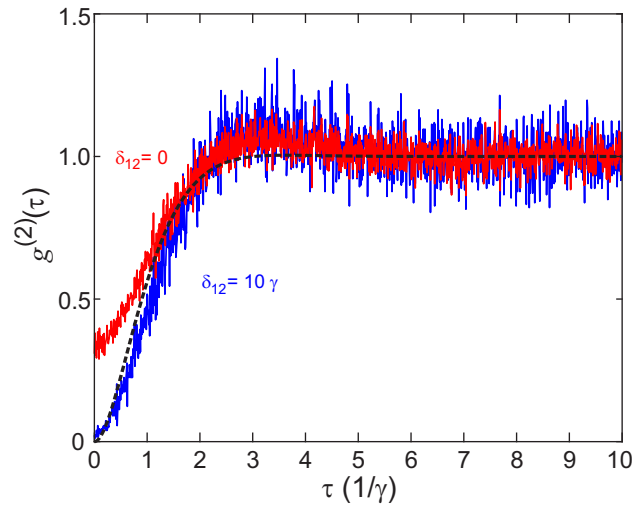}}}
	\caption{The $g^{(2)}$-function including interactions between the atoms is plotted for interaction strength $\delta_{12}=0$ (red curve) and $\delta_{12}=10~\gamma$ (blue curve). The length of the simulated trajectories is $T_1=2\times 10^5/\gamma$ for the one-atom case and $T_2=1\times 10^5/\gamma$ for the two atom case, with $\gamma=\gamma_1=\gamma_2$. The resolution is set to a value of $dt=10^{-3}/\gamma$. We checked that the simulation result is independent from the exact value of $dt$. For the one-atom case the Rabi frequencies are chosen as $\Omega_1=\gamma$ and $\Omega_2=0$, whereas for the two-atom case $\Omega_1=\Omega_2=\gamma$. The detunings are chosen to be $\Delta=0$. For comparison, we included the analytical $g^{(2)}$-function of the light emitted from a single emitter (black dashed line) with $\Omega=\gamma$.}
	\label{fig:g2_ia}
\end{figure}
\section{Quantum jump approach including dipole-dipole interactions}
\begin{figure}[ht]
	\centerline{\scalebox{.9}{\includegraphics{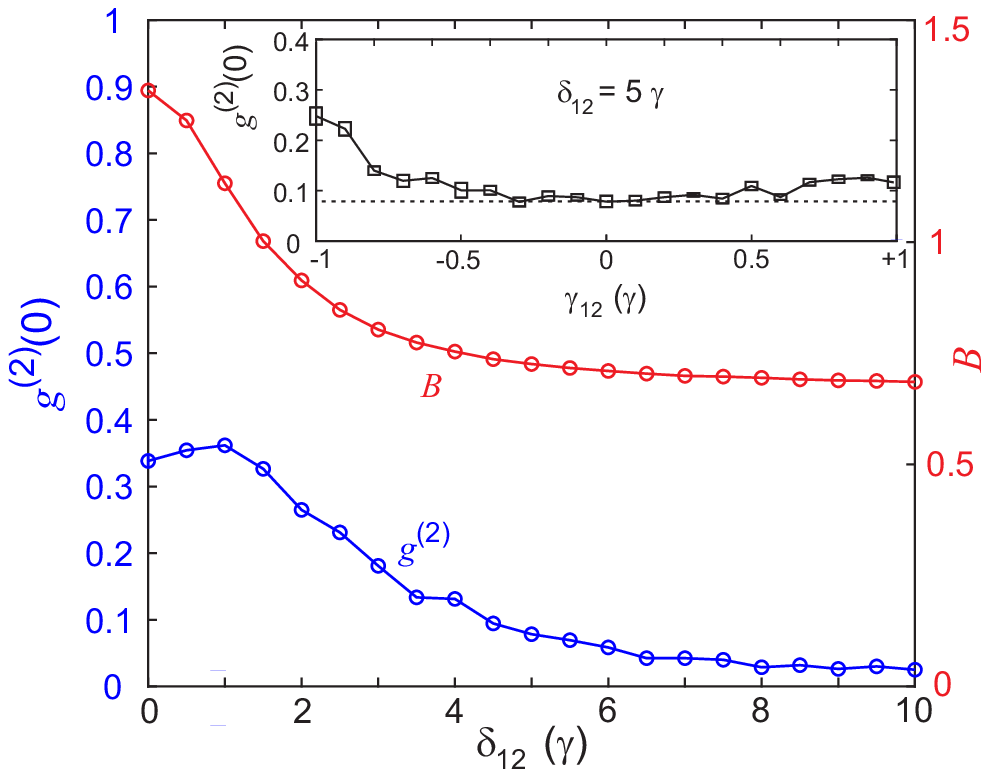}}}
	\caption{The value of $g^{(2)}(0)$ (left axis, blue line) tends to zero for increasing interaction strength $\delta_{12}$, whereas the brightness $B$ (right axis, red line) saturates at a value of $B=2/3$ (for $\mu=1$). In the main part of the figure, the collective correction is set to zero, i.e. $\gamma_{12}=0$. Collective decay can however increase the value of $g^{(2)}(0)$, as shown in the inset for an intermediate value of $\delta_{12}=5\gamma$. This effect disappears for large interaction strength. For small values of $\delta_{12}$ it is more pronounced and can even lead to the bunching of emitted photons.}
	\label{fig:g2_delta}
\end{figure}
In this section we analyze general situations which do not depend on a specific geometry, and in particular not on the presence of a nanotip. For that reason we regard the interaction strength $\delta_{12}$ as free parameter. Moreover, we set both decay rates to the same value $\gamma=\gamma_1=\gamma_2$, and -- first -- neglect the collective decay by setting $\gamma_{12}=0$. The latter assumption is justified in the sense that in real situations $\delta_{12}$ can have arbitrary large values whereas the collective correction term is always limited to the range $\gamma_{12}\in \left[-\gamma,\gamma\right]$. Time trajectories of the Hamiltonian (\ref{eq:H_tot}) including the interaction terms are simulated using the quantum jump approach explained in section \ref{sec:QJA}. In order to take the Poisson-distribution of atom numbers into account, the relative length of the corresponding trajectories is adjusted accordingly. We chose here an average atom number of $\mu=1$. With $P_1(1)=0.37$ and $P_1(2)=0.18$ the trajectory of the one-atom case is twice as long as that of the two atom case. The no-atom case with $P_1(0)=0.37$ is not considered as it does only reduce the brightness, but leaves the $g^{(2)}$-function unchanged. Cases with more than two atoms are also not considered, because their contribution is small. Each simulation results in a series of time-stamps when photons are emitted corresponding to the times a jump occurs in the simulation. Finally, the time trajectories for the one-atom and two-atom case are appended to each other, and from the resulting trajectory the $g^{(2)}$-function is calculated. Corresponding $g^{(2)}$-functions derived from such simulated time trajectories are exemplarily shown in Fig.~\ref{fig:g2_ia}. From time trajectories as shown in Fig.~\ref{fig:g2_ia} the value of $g^{(2)}(0)$ is evaluated and plotted in Fig.~\ref{fig:g2_delta}. Without interactions $g^{(2)}(0)=0.34$ coincides in good approximation with the value of a Poissonian statistical mixture of $N$ atoms as derived in Eq. (\ref{eq:g2_N_mean}). This justifies the validity of the approximation of neglecting emission from more than two atoms for the present case of $\mu=1$. For increasing interaction strength the value of $g^{(2)}(0)$ tends to zero, and the emission resembles more and more that of a single atom. The brightness however does not tend to zero, but saturates at a value given by the fraction of the duration of the single-atom trajectory as compared to the total time. This corresponds to the fact that single photons are emitted with the single-emitter rate whenever a single atom is in the volume whereas at most one photon is emitted at times when two atoms are in the volume. In order to take the effect of collective decay into account, which was neglected in the main part of Fig.~\ref{fig:g2_delta}, we simulated time trajectories for fixed values of $\delta_{12}$ and varying $\gamma_{12}$. The result is plotted in the inset of Fig.~\ref{fig:g2_delta} for $\delta_{12}=5\gamma$. We observe that the value of  $g^{(2)}(0)$ is increased for $\gamma_{12}\rightarrow \pm \gamma$, the effect being more pronounced for small interaction strength. For large values of $\delta_{12}$ the influence of $\gamma_{12}$ on the $g^{(2)}$-function disappears. Concluding, photon antibunching can be reached in a statistical mixture of $N$ interacting atoms without the drawback of a low brightness which is a known issue in standard non-deterministic single photon sources. The following section analyzes the specific situation of atoms that are excited by the light field emerging from a tapered nanotip as sketched in Fig.~\ref{fig:tip_g2}. Collective dispersive and radiative effects of the two atoms as well as the dispersive and radiative influence of the nanosphere on the individual atoms are taken explicitely into account.

\section{Cooperative effects of two atoms at the nanotip}
\begin{figure}[ht]
	\centerline{\scalebox{.85}{\includegraphics{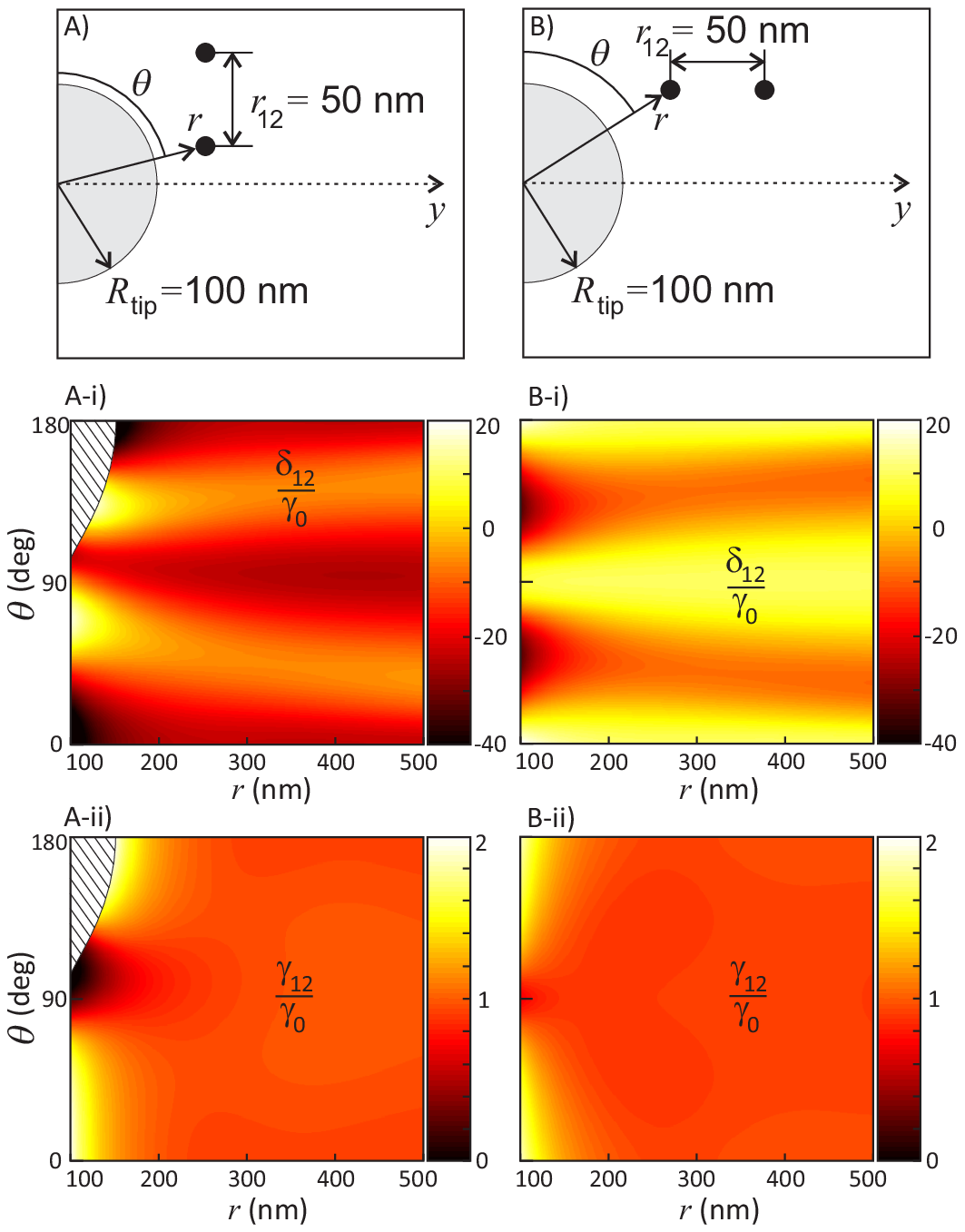}}}
	\caption{Two atoms are positioned at a fixed distance $r_{12}=50~\mathrm{nm}$ in $z$-direction (geometry A), resp. in $y$-direction (geometry B). In both cases the collective correction $\gamma_{12}$ and the dipole-dipole interaction strength $\delta_{12}$ are calculated as function of the atomic positions defined by $r$ and $\theta$ for $\varphi=90^\circ$, relative to the sphere. Note that geometry A is asymmetric with respect to the $y$-axis. The white dashed region in the upper left corner of A-i) and A-ii) thus corresponds to the situation where one of the atoms would be situated within the sphere.}
	\label{fig:gamma_delta_tip}
\end{figure}

\begin{figure}[ht]
 	\centerline{\scalebox{.9}{\includegraphics{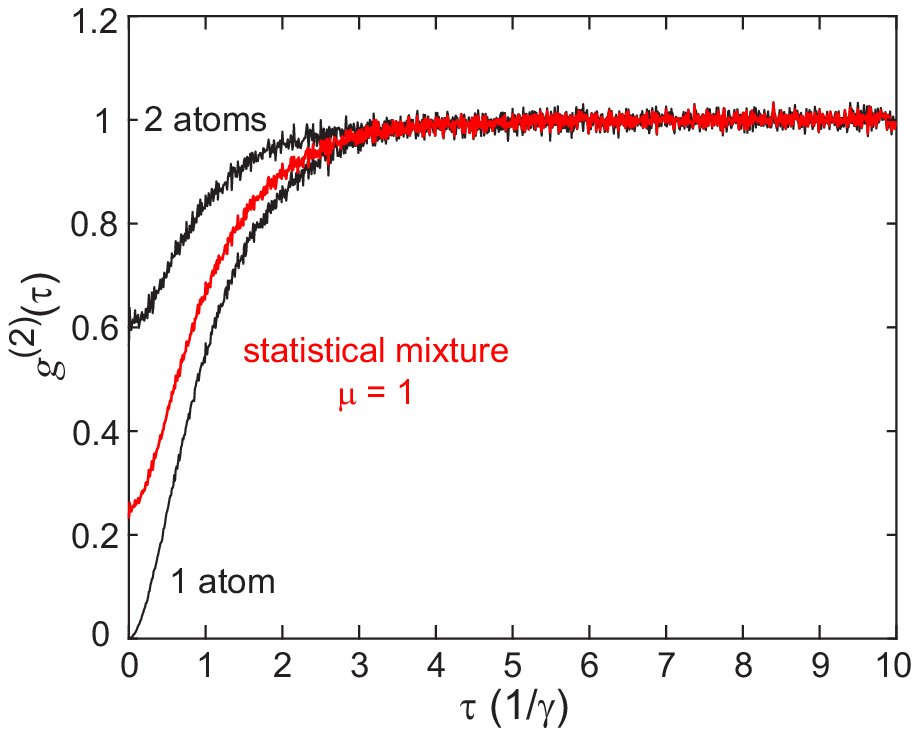}}}
 	\caption{Autocorrelation $g^{(2)}$-function of photons emitted by atoms in the volume close to the nanotip. The countrate with two atoms in the volume is reduced by dipole-dipole interactions. Thus,  the averaged $g^{(2)}$-function exhibits improved antibunching. For the 1-atom (2-atoms) case 100 (50) realizations are simulated, where the length of each simulated trajectory is $T_1=1\times 10^5/\gamma_0$, and the resolution is $dt=10^{-3}/\gamma_0$. }
 	\label{fig:g2_tip}
 \end{figure}

We consider two quantum emitters of transition frequency $\omega_0=k_0c$ located at positions $\mathbf{r}_1=r_1\hat{\mathbf{r}}_1$ and $\mathbf{r}_2=r_2\hat{\mathbf{r}}_2$ in the vicinity of a spherical nanotip of radius $R_\mathrm{tip}$ ($r_1,r_2>R_\mathrm{tip}$), dielectric permittivity $\varepsilon$, and centered at the origin.
The interatomic separation is $r_{12}=|\mathbf{r}_1-\mathbf{r}_2|$, as shown in Fig.~\ref{fig:gamma_delta_tip}A) and B).
For large $r_{12}$, i.e., $k_0r_{12}\gg1$, the collective correction to the decay rate and dipole-dipole interaction strength are negligible for any mutual orientation of the dipole moments, $\gamma_{12}=\delta_{12}=0$ . Conversely, in the limit of separations $r_{12}$ much smaller than the optical wavelength, i.e. $k_0r_{12}\ll1$, the collective correction and the dipole-dipole interaction strength are given by the approximate expressions 
\begin{equation}
\gamma_{12} \approx \sqrt{\gamma_1\gamma_2}(\hat{\mathbf{d}}_1\cdot\hat{\mathbf{d}}_2)~,
\label{eq:gamma12_approx} 
\end{equation}
and 
\begin{equation}
\delta_{12} \approx 3\sqrt{\gamma_1\gamma_2}[(\hat{\mathbf{d}}_1\cdot\hat{\mathbf{d}}_2) - 3(\hat{\mathbf{d}}_1\cdot\hat{\mathbf{r}}_{12})(\hat{\mathbf{d}}_2\cdot\hat{\mathbf{r}}_{12})]/4(k_0r_{12})^3~,
\label{eq:delta12_approx} 
\end{equation}
respectively, where $\mathbf{d}_j=d_{eg}\hat{\mathbf{d}}_j$ and $\mathbf{r}_{12}=r_{12}\hat{\mathbf{r}}_{12}$ with the dipole matrix element $d_{eg}$~\cite{Agarwal_PhysRevA45_1992,Agarwal_PhysRevA57_1998,Akram_PhysRevA62_2000,Ficek_PhysRep372_2002}.
Our simulations are based on the general expression valid for arbitrary separations $r_{12}$~\cite{Akram_PhysRevA62_2000,Ficek_PhysRep372_2002}
\begin{align}
\frac{\gamma_{12}}{\sqrt{\gamma_1\gamma_2}} &= \frac{3}{2}\bigg\{\left[\hat{\mathbf{d}}_1\cdot\hat{\mathbf{d}}_2 - (\hat{\mathbf{d}}_1\cdot\hat{\mathbf{r}}_{12})(\hat{\mathbf{d}}_2\cdot\hat{\mathbf{r}}_{12})\right]\frac{\sin(k_0r_{12})}{k_0r_{12}}\nonumber\\
&+\left[(\hat{\mathbf{d}}_1\cdot\hat{\mathbf{d}}_2) - 3(\hat{\mathbf{d}}_1\cdot\hat{\mathbf{r}}_{12})(\hat{\mathbf{d}}_2\cdot\hat{\mathbf{r}}_{12})\right]\nonumber\\
&\times\left[\frac{\cos(k_0r_{12})}{(k_0r_{12})^2}-\frac{\sin(k_0r_{12})}{(k_0r_{12})^3}\right]\bigg\},\label{gamma-ij}\\
\frac{\delta_{12}}{\sqrt{\gamma_1\gamma_2}} &= \frac{3}{4}\bigg\{-\left[\hat{\mathbf{d}}_1\cdot\hat{\mathbf{d}}_2 - (\hat{\mathbf{d}}_1\cdot\hat{\mathbf{r}}_{12})(\hat{\mathbf{d}}_2\cdot\hat{\mathbf{r}}_{12})\right]\frac{\cos(k_0r_{12})}{k_0r_{12}}\nonumber\\
&+\left[(\hat{\mathbf{d}}_1\cdot\hat{\mathbf{d}}_2) - 3(\hat{\mathbf{d}}_1\cdot\hat{\mathbf{r}}_{12})(\hat{\mathbf{d}}_2\cdot\hat{\mathbf{r}}_{12})\right]\nonumber\\
&\times\left[\frac{\sin(k_0r_{12})}{(k_0r_{12})^2}+\frac{\cos(k_0r_{12})}{(k_0r_{12})^3}\right]\bigg\}~.\label{delta-ij}
\end{align}
Equations~(\ref{gamma-ij}) and (\ref{delta-ij}) are well-known expressions for the collective correction to the decay rate and dipole-dipole interaction strength of two dipole emitters in vacuum \cite{Ficek_Book_2005}.
However, here we are interested in the influence of a spherical nanotip on the two-atom dynamics, which is expected to be modified not only by the collective damping but also by the Purcell effect~\cite{Zadkov_PhysRevA85_2012}. To this end, we use Eqs.~(\ref{gamma-ij}) and (\ref{delta-ij}) and substitute the free-space decay rates by the decay rates $\gamma_1$ and $\gamma_2$ of the corresponding atoms not in free space but in the vicinity of the spherical nanoparticle. We have verified that this approximation provides accurate results when compared with the exact solution as long as we consider moderate permittivities with ${\rm Im}(\varepsilon)=0$ and $\varepsilon>0$, i.e., there is no ohmic losses on the dielectric nanotip. To simplify our discussion, we also consider that the spherical nanotip is much smaller than the emission wavelength, i.e., $k_0R_{\mathrm{tip}}\ll1$.
Under this last assumption, the single-atom decay rate is given by~\cite{Carminati_OptComm261_2006}
\begin{align}
\gamma_{j} = \gamma_j^{\perp} (\hat{\mathbf{d}}_j\cdot\hat{\mathbf{r}}_j)^2 + \gamma_j^{||}\left[1-(\hat{\mathbf{d}}_j\cdot\hat{\mathbf{r}}_j)^2\right],
\end{align}
where
\begin{align}
\frac{\gamma_j^{\perp}}{\gamma_{0}} \approx &1 + \frac{3k_0^3}{2\pi}{\rm Im}\bigg\{\alpha(\omega_0)e^{2ik_0r_j}\bigg[-\frac{1}{(k_0r_j)^4}+\frac{2}{i(k_0r_j)^5}\nonumber\\
&+\frac{1}{(k_0r_j)^6}\bigg]\bigg\},\label{gamma-perp}\\
\frac{\gamma_j^{||}}{\gamma_{0}} \approx & 1 + \frac{3k_0^3}{8\pi}{\rm Im}\bigg\{\alpha(\omega_0)e^{2ik_0r_j}\bigg[\frac{1}{(k_0r_j)^2}-\frac{2}{i(k_0r_j)^3}\nonumber\\
&-\frac{3}{(k_0r_j)^4}+\frac{2}{i(k_0r_j)^5}+\frac{1}{(k_0r_j)^6}\bigg]\bigg\},\label{gamma-para}
\end{align}
with $\gamma_{0}=d_{eg}^2\omega_0^3/3\pi\varepsilon_0\hbar c^3$ the Einstein $A$ coefficient.
We have verified that Eqs.~(\ref{gamma-perp}) and (\ref{gamma-para}) are very good approximations for the exact expressions in \cite{Chew_JCPhys87_1987} when the sphere is subwavelength. The polarizability $\alpha(\omega)$ of a subwavelength nanosphere, accounting for radiation damping, is~\cite{Carminati_OptComm261_2006}
\begin{align}
\alpha(\omega)  =\frac{{\alpha}_0}{1-i(k^3/6\pi){\alpha}_0},\label{alpha}
\end{align}
where ${\alpha}_0=4\pi R_\mathrm{tip}^3(\varepsilon-1)/(\varepsilon+2)$ is the quasi-static polarizability of the nanosphere. The transmitted/scattered electric field by the spherical nanotip determines the dipole moment orientation. Within the fiber, we consider a $z$-polarized electromagnetic plane wave with wave vector $\mathbf{k}_L=k_L\hat{\mathbf{y}}$ and amplitude $E_0$, where $\omega_L=k_Lc$ is the laser frequency. Since we work with light tuned close to the atomic resonance, we use $k_L\cong k_0$. Outside the spherical nanotip, at the position $\mathbf{r}=(r,\theta,\varphi)$ the local electric field is given by Ref.~\cite{Jackson_Book}, p. 411 as 
\begin{align}
\mathbf{E}(\mathbf{r},\omega_L)&=E_0\frac{{\alpha}(\omega_L)}{4\pi r^3}
e^{ik_0r}\bigg\{2\cos\theta\left(1-ik_0r\right)\hat{\mathbf{r}}\nonumber\\
&+\sin\theta\left[1-ik_0r-(k_0r)^2\right]\hat{\boldsymbol{\theta}}\bigg\},\label{E-field}
\end{align}
where ${\alpha}(\omega_L)$ is the effective polarizability of the spherical nanotip given in Eq.~(\ref{alpha}). Here, the near- and far-field contributions are proportional to $1/r^3$ and $k_0^2/r$, respectively, whereas the intermediate region are mainly governed by the term proportional to $i k_0/r^2$. Note that the electric field fixes the spherical coordinate system. In this case, the $y$ axis is directed along the fiber and parallel to the horizontal axis of Fig.\ref{fig:tip_g2}. This choice of coordinates implies $E_{\varphi}=0$, which simplifies the expressions. Finally, we consider that the dipole moment of atom $j$ is directed along the local electric field at $\mathbf{r}_j$~\cite{Zadkov_PhysRevA85_2012}, i.e., $\mathbf{d}_j=d_{eg}\{\mathbf{E}(\mathbf{r}_j,\omega_L)\}/||\{\mathbf{E}(\mathbf{r}_j,\omega_L)\}||$.
This leads to the Rabi frequency
\begin{align}
\Omega_j(\mathbf{r}_j,\omega_L)=\frac{d_{eg}}{\hbar}||\left\{\mathbf{E}(\mathbf{r}_j,\omega_L)\right\}||,\label{Rabi}
\end{align}
where the electric field is given in Eq.~(\ref{E-field}). Note that we drop the time-harmonic dependence $e^{-\imath\omega_L t}$ in Eq.~(\ref{E-field}), so that Eq.~(\ref{Rabi}) is the usual definition of the Rabi frequency. 
Two specific geometries are considered in Fig.~\ref{fig:gamma_delta_tip}A) and B). In these geometries the interatomic distance is kept fixed at $r_{12}=50~\mathrm{nm}$, and the position of the atoms relative to the tip is varied. The two geometries correspond to situations where the interatomic vector  $\hat{\mathbf{r}}_{12}$ is parallel to the y- and z-axis, respectively. The results for the collective correction and the dipole-dipole interaction strength are plotted in Figs.~\ref{fig:gamma_delta_tip}(A/B-i/ii). The interaction strength is large, i.e. $\left|\delta_{12}\right|\gg\gamma_{0}$, in a large region around the tip, such that a reduction of the value of $g^{(2)}(0)$ can be expected according to Fig.~\ref{fig:g2_ia}.\\
In order to predict the result of a real experiment we average over many arbitrarily chosen atomic positions. For this purpose atoms are randomly placed around the tip in a half-sphere shell in 3D with inner radius given by the tip size, $R_i=R_\mathrm{tip}=100~\mathrm{nm}$, and outer radius twice as large, $R_o=200~\mathrm{nm}$. Its volume is $V=2.9\times 10^{-14}\mathrm{cm}^3$. Thus, a mean number of $\mu=1$ atom in the volume corresponds to an atomic density of $\rho=3.4\times 10^{13}\mathrm{cm}^{-3}$ which can be reached in experiments with ultracold atoms. In order to take into account the poissonian distribution of atoms in the volume for $\mu=1$ where the probability of having one atom in the volume is twice as large as having two atoms in the volume, twice as many realizations are simulated for the 1-atom case than for the 2-atom case. We simulate 100 and 50 realizations, respectively. For each realization a quantum Monte Carlo simulation is carried out with the respective parameters $\gamma_1$, $\gamma_2$, $\gamma_{12}$, $\delta_{12}$, $\Omega_1$, and $\Omega_2$ given by the positions of the atoms relative to the tip. The Rabi frequencies are an open parameter, because they are proportional to the laser power used in the experiment. In the simulation they are scaled such that a Rabi frequency of $\Omega=\gamma_0$ is reached at the surface of the nanotip, i.e. at the position $r=100~\mathrm{nm}$, $\theta=90^\circ$, and $\varphi=90^\circ$ within the $yz$-plane. Moreover, in the 1-atom case, the Rabi frequency acting on the second atom and the collective parameters are artificially set to zero, $\Omega_2=0$, $\delta_{12}=0$, and $\gamma_{12}=0$. Thus, the second atom is not excited and has got no influence on the first atom. A $g^{(2)}$-function is calculated from the trajectory of each QMC-simulation. The individual $g^{(2)}$-functions are weighted with their corresponding rate of photon emission and are averaged. Such averages are calculated for the 1-atom cases and the 2-atom cases, and for the statistical mixture of both. The results are plotted in Fig.~\ref{fig:g2_tip}. The averaged $g^{(2)}$-function of the 1-atom cases drops to zero for $\tau\rightarrow0$, as expected for a single emitter. In the 2-atom cases the value of the averaged $g^{(2)}(0)>0.5$. The dipole-dipole interaction of both atoms leads on average to an increase of $g^{(2)}(0)$ above the non-interacting case and thus to bunching, although values of $g^{(2)}(0)\ll 0.5$ (i.e. strong antibunching) can be reached for specific positions of the two atoms. However, the photon emission rate of the 2-atom case is reduced, compared to the non-interacting situation. The averaged photon emission rates in Fig.~\ref{fig:g2_tip} are given in the 1-atom case by $R_\mathrm{1A}=0.30~\gamma_{0}$, in the 2-atom case by  $R_\mathrm{2A}=0.42~\gamma_{0}$, and in the statistical mixture by $R_\mathrm{mix}=0.34~\gamma_{0}$. Without interactions the emission rate of the 2-atom case would be $R_\mathrm{2A}=2 R_\mathrm{1A}=0.6~\gamma_{0}$. Thus, the value of $g^{(2)}(0)$ of the statistical mixture of atoms is reduced to a value of $g^{(2)}(0)=0.25$ smaller than without interactions. At the same time, the brightness $B$ defined by (\ref{eq:brightness}) is with $B=\frac{R_\mathrm{mix}}{R_\mathrm{1A}}=1.13$ comparable to the 1-atom case. 

\section{Conclusion and outlook} 
The paper investigates the light statistics of the fluorescence from statistical mixtures of one and two atomic emitters within a nanoscale volume. The $g^{(2)}$-function of the emitted light is derived by a Quantum Monte Carlo approach including dipole-dipole interactions between the atoms and collective decay. The main result of the paper is the fact that these interactions can improve the antibunching while the brightness of the source tends to a constant value. This result is interesting for the design of novel single photon sources that are not based on single emitters but on statistical ensembles. As application, we investigate the situation of cold atoms in a nanoscale volume at the apex of a nanofiber tip and predict a reduction of the $g^{(2)}$-function due to atomic interactions to a value of $g^{(2)}(0)=0.25$. We expect that the quality of antibunching can be further improved by optimizing the radius of the nanotip, the atomic density and the light power used. Furthermore, manipulating the shape of the optical near-field, for instance by depositing tailored plasmonic nanostructureson the tip, may be a method to restrict the excitation of atoms to those positions where the antibunching is more pronounced. Moreover, the fluorescence may be also collected by the nanotip which puts further restrictions on the atomic positions that contribute to the detected signal \cite{Auwaerter13}. As a final remark we would like to add that the method we introduce can be used also for other types of interactions, for instance van-der Waals interactions between Rydberg atoms \cite{Beguin13}.

\section{Acknowledgement}
This work was performed in the context of the European COST Action MP1302 Nanospectroscopy. S.S. acknowledges support by the Fulbright-Cottrell Award. E. S. acknowledges support by the ColOpt - EU H2020 ITN 721465. T.J.A., R.B., and Ph.W.C. acknowledge FAPESP (Grant Nos. 2015/21194-3, 2014/01491-0, and 2013/04162-5, respectively) for financial support.

\section{Competing financial interests}
Competing financial interests do not exist.

\end{document}